\documentclass[conference]{IEEEtran}

\usepackage{booktabs} % For formal tables
\usepackage{cite}
\usepackage[TABBOTCAP]{subfigure}
\usepackage{tabularx}
\usepackage{graphicx}
\usepackage{color}
\usepackage{xspace}
\usepackage{thumbpdf}
\usepackage{listings}
\usepackage{verbatim}
\usepackage{amsmath}
\usepackage[linesnumbered,ruled,vlined]{algorithm2e}
\usepackage{balance}

\usepackage{hyperref}
\hypersetup{
    colorlinks=true,
    linkcolor=blue,          % color of internal links (change box color with linkbordercolor)
    citecolor=magenta,        % color of links to bibliography
    filecolor=blue,      % color of file links
    urlcolor=red           % color of external links}
}

\usepackage{multirow}

\usepackage{array}
\newcolumntype{P}[1]{>{\centering\arraybackslash}p{#1}}

\usepackage{booktabs}
\usepackage{colortbl}

\usepackage{url}
\usepackage{float}
\usepackage{amssymb}
\usepackage{pifont}

\newcommand{\sys}{{\textsc{P4COM}}\xspace}

\newcolumntype{Y}{>{\centering\arraybackslash}X}

\usepackage{array}
\newcolumntype{L}[1]{>{\raggedright\let\newline\\\arraybackslash\hspace{0pt}}m{#1}}
\newcolumntype{C}[1]{>{\centering\let\newline\\\arraybackslash\hspace{0pt}}m{#1}}
\newcolumntype{R}[1]{>{\raggedleft\let\newline\\\arraybackslash\hspace{0pt}}m{#1}}

\usepackage{amssymb}% http://ctan.org/pkg/amssymb
\usepackage{pifont}% http://ctan.org/pkg/pifont
\newcommand{\cmark}{\ding{51}}%
\newcommand{\xmark}{\ding{55}}%

\begin{document}

%make title bold and 14 pt font (Latex default is non-bold, 16 pt)
\title{P4COM: In-Network Computation with Programmable Switches}

%for single author (just remove % characters)
\author{\IEEEauthorblockN{Ge~Chen$^1$, Gaoxiong~Zeng$^2$, Li~Chen$^3$} \IEEEauthorblockA{\textit{$^1$Alibaba~~~~~~$^2$HKUST~~~~~~ $^3$Huawei}}}
% copy the following lines to add more authors
% \and
% {\rm Name}\\
%Name Institution
%} % end author

% \renewcommand{\shortauthors}{G. Zeng et al.}
% \renewcommand{\shorttitle}{Leverage Programmable Data Plane to fuel \\ Distributed Applications}

%%
%% The code below should be generated by the tool at
%% http://dl.acm.org/ccs.cfm
%\begin{CCSXML}
%<ccs2012>
%<concept>
%<concept_id>10003033.10003039.10003048</concept_id>
%<concept_desc>Networks~Transport protocols</concept_desc>
%<concept_significance>500</concept_significance>
%</concept>
%</ccs2012>
%\end{CCSXML}
%
%\ccsdesc[500]{Networks~Transport protocols}

%\keywords{Programmable Dataplane, Distributed Application}

\maketitle
\thispagestyle{plain}
\pagestyle{plain}

\begin{abstract}
Traditionally, switches only provide forwarding services and have no credits on computation in distributed computing frameworks. The emerging programmable switches make in-network computing (INC) possible, i.e., offloading some computation to the switch data plane. While some proposals have attempted to offload computation onto special hardwares (e.g., NetFPGA), many practical issues have not been addressed. Therefore, we propose P4COM---a user-friendly, memory-efficient, and fault-tolerant framework realizing in-network computation (e.g., MapReduce) with programmable switches.

P4COM consists of three modules. First, P4COM automatically translates application logic to switch data plane programs with a lightweight interpreter. Second, P4COM adopts a memory management policy to efficiently utilize the limited switch on-chip memory. Third, P4COM provides a cutting-payload mechanism to handle packet losses. We have built a P4COM prototype with a Barefoot Tofino switch and multiple commodity servers. Through a combination of testbed experiments and large-scale simulations, we show that P4COM is able to achieve line-rate processing at 10Gbps links, and can increase the data shuffling throughput by 2--5$\times$ for the MapReduce-style applications.
\end{abstract}

\section{Introduction}\label{sec:intro}

Internet services, such as web search~\cite{singh2015jupiter}, social networking~\cite{roy2015inside}, and e-commerce~\cite{amazon}, have become prevalent since last decade. To scale out, many of these services leverage distributed computation frameworks like MapReduce~\cite{mr2004dean, hadoop, spark, spark2012matei} to process data partitions simultaneously on many machines and aggregate the results by network communication. This partition-aggregation pattern of MapReduce frameworks, however, introduces significant burden on network
communication.  For example, the aggregation traffic accounts for 46\% of the total network traffic in Facebook's data centers~\cite{lef2013mosharaf}. 

The aggregation traffic in data centers have become severe bottlenecks for the following two reasons. First, for intra-rack aggregation, the inbound bandwidth (e.g., 10Gbps~\cite{singh2015jupiter}) at the receiver side is significantly smaller than the sum of the outbound bandwidth from the senders. Second, for inter-rack communication, data center networks often adopt a certain degree of over-subscription \cite{vl2009greenberg, traffic2010benson}, making inter-rack bandwidth far smaller than intra-rack bandwidth.

To overcome the network bottleneck, some work~\cite{varys2014mosharaf, bai2015pias, coda2016hong, karuna, bytescheduler, dlcp} has proposed mechanisms to schedule the aggregation traffic according to application-level priorities to better utilize the inbound bandwidth at the receiver side. However, these approaches do not directly reduce the traffic volume and may still suffer from low throughput. Other work~\cite{al2008fattree, vl2009greenberg, osa, netagg2014mai} has proposed special network topologies and middleboxes to alleviate the unbalance between outbound and inbound bandwidth. While these solutions can alleviate the bandwidth bottleneck from data aggregation, they are cost-prohibitive for real deployment due to high bandwidth over-provisioning.

To address the problem, inspired by the emerging programmable switches, we ask the following question: \textit{can we offload some computation tasks to switches in a reliable manner such that the aggregation traffic can be reduced?}
We answer this question affirmatively in this paper.

\begin{figure*}[ht]
\centering
    \subfigure[Conventional MapReduce]{
    	\includegraphics[scale=0.5]{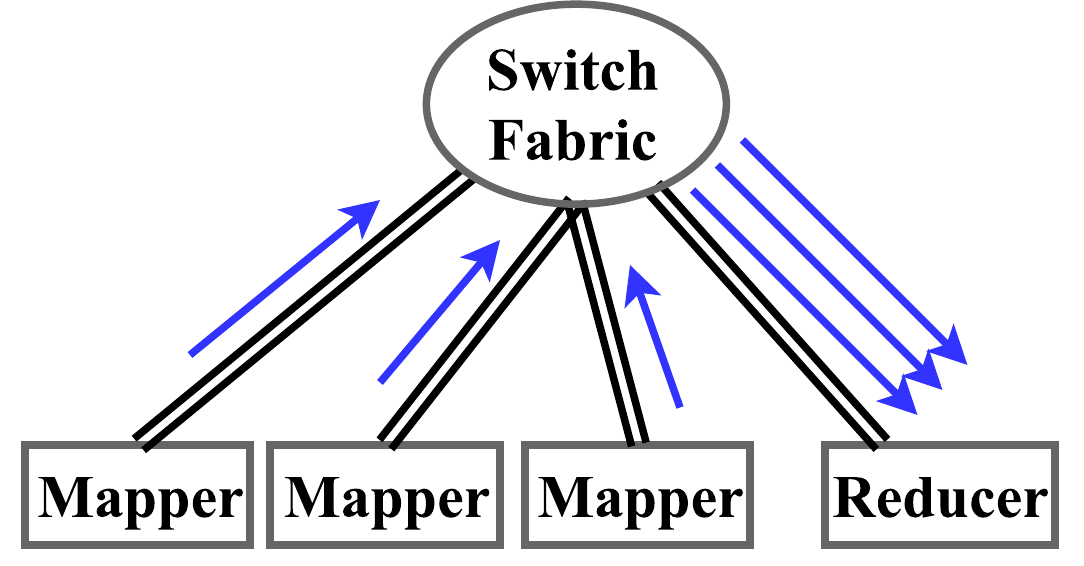}
    }
    \subfigure[Middlebox aggregation (e.g., NetAgg)]{
    	\includegraphics[scale=0.5]{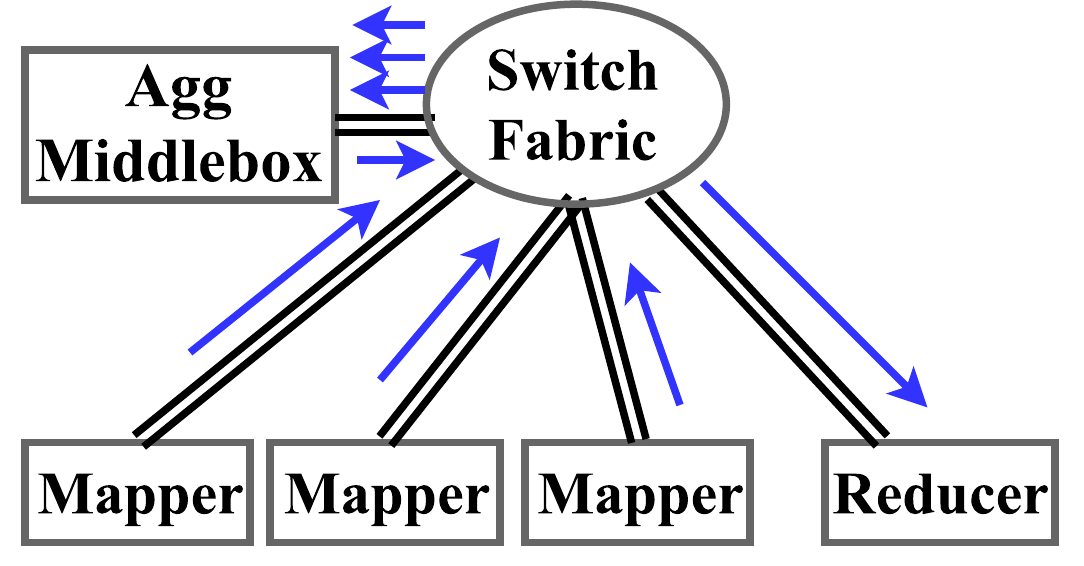}
    }
    \subfigure[In-network computation (e.g., P4COM)]{
    	\includegraphics[scale=0.5]{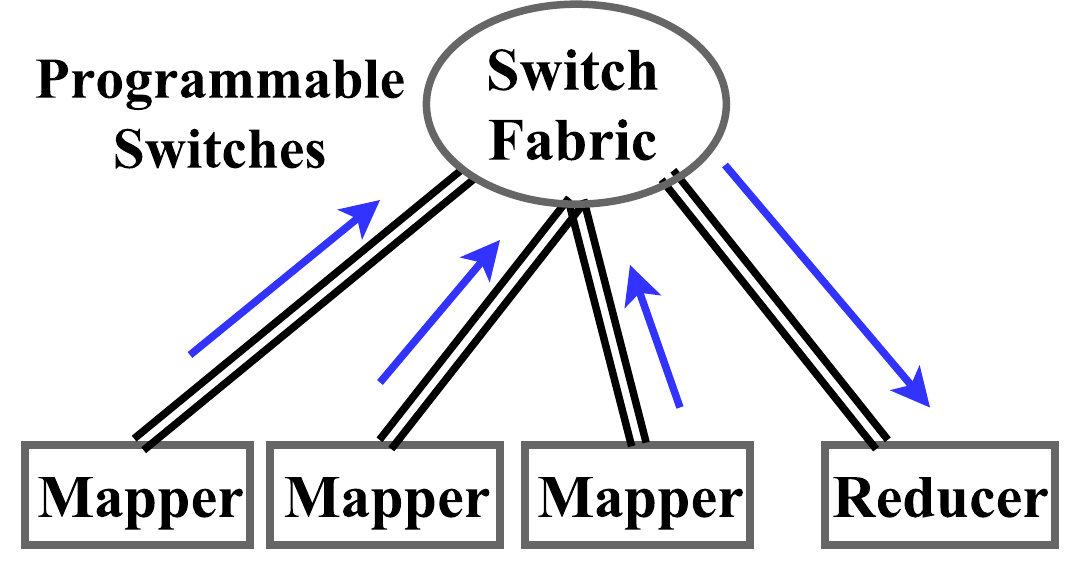}
	}
    \caption{Data communication for conventional MapReduce, via middlebox aggregation (e.g., NetAgg), or with in-network computation (e.g., P4COM).}
    \label{fig:motivation}
    \vspace{-1em}
\end{figure*}

Traditionally, switches only provide data forwarding services, while all computation tasks are performed on servers. The emerging programmable switches make in-network computation (INC) possible, potentially reducing the aggregation communication bottleneck.
Figure~\ref{fig:motivation} illustrates the promising performance of in-network computation. 
Conventionally, application data is transparent to switches, and all shuffling data is forwarded to the dedicated reducer. The middlebox solutions such as NetAgg~\cite{netagg2014mai} redirect application-specific traffic to middleboxes, and data is aggregated before resending to destination. With in-network computation, the bottleneck switch fabric caches partial shuffling data, and performs computation in the data plane before forwarding to reducers.

However, offloading computation tasks to switches is non-trivial and has the following challenges. First, the native switch interfaces (CLI and APIs) are primarily designed for network administrators to do network configurations and monitoring, which are not friendly for application developers to program computation tasks. Second, switches have limited on-chip memory for flow tables and packet buffers, while many computation tasks require a significant amount of memory. By performing computation tasks on switches, the precious switch memory resources would soon become a bottleneck. Third, aggregation traffic may suffer from packet loss in the network. If not being correctly handled , the packet loss would cause incorrect computation results in the switches.

In this paper, we propose P4COM, an end-to-end framework to address these challenges and enable computation tasks to be offloaded to programmable switches. The key idea of P4COM is a \emph{reliable} packet processing pipeline that can parse, extract, compute and store application-specific data.
To simplify the programming for application developers, we develop an interpreter for P4COM to translate MapReduce logics into P4 programs.
To handle the limited switch on-chip memory, we design a memory management module for P4COM to avoid switch memory overflow while keeping high usage ratio.
To achieve fault tolerance, we apply an improved cutting-payload transport protocol to quickly recover from packet loss and avoid errors on duplicate packet retransmissions. 
Experimental results show that P4COM can increase shuffling throughput by 2--5$\times$ and achieve near line-rate goodput in many-to-one mapper-reducer scenarios.

The contributions of this paper are summarized as follows.
\begin{itemize}
    \item We propose P4COM, a general computation framework that leverages the capability and flexibility of programmable switches to accelerate distributed applications and reduce shuffling traffic in the network.
    \item We evaluate P4COM on multiple benchmark datasets via simulation, and demonstrate the advantages of our memory management and fault tolerance design.
    \item We build a P4COM prototype on a small-scale testbed of a Barefoot Tofino switch and several commodity servers, and validate the superior performance of P4COM.
\end{itemize}

The paper is organized as follows. First, we introduce the MapReduce applications and the recent progress on programmable switches, and motivate the idea of in-network computation (\S\ref{sec:background}).
Then, we present the design of P4COM (\S\ref{sec:design}), and explain the use cases and limitations of our current framework (\S\ref{sec:discussion}).
Finally, we conduct both simulation experiments and testbed measurements (\S\ref{sec:evaluation}) to demonstrate the superior performance by offloading computation tasks to programmable switches.
\section{Background and Motivation} \label{sec:background}

In this section, we first give an introduction on the workflow of
MapReduce-based frameworks (\S\ref{subsec:mr}) and the recent development of programmable switches (\S\ref{subsec:prog_switch}). Then, we motivate the idea of in-network computation (\S\ref{subsec:motivation}).

\subsection{MapReduce} \label{subsec:mr}
MapReduce~\cite{mr2004dean} is a programming model that is primarily designed to process large datasets on a computer cluster. The datasets are usually stored in a filesystem or in a database within the cluster. A general MapReduce workflow consists of three steps, namely a ``Map'' step, a ``Shuffle'' step and a ``Reduce'' step. These steps are running in sequence after the MapReduce job is launched. First in the ``Map'' step, each worker node or process (called ``mapper'') executes a ``map()'' function on its local data partition and the data is stored temporarily. Second, a ``Shuffle'' step distributes the intermediate data generated by mappers to reducers (worker node or process that executes reduce() function). Finally, a ``Reduce'' step processes received data on each reducer in parallel, and the final results are returned to the user. The basic data structure of MapReduce is the pair of key-value. The ``map()'' function specifies how to parse data source to key-value pairs and the ``reduce()'' function defines the computation on values for the same key.

Several frameworks~\cite{hadoop, spark, spark2012matei} have been proposed to do computation tasks based on the MapReduce model. In general, these frameworks assign one node as a master and the others as workers. The role of the master node is to monitor the MapReduce process, manage resources, and assign mappers and reducers to different worker nodes. A principle in task allocation is to put mappers close to data locations. Since the mappers and reducers are distributed randomly in the cluster, the ``Shuffle'' step will generate traffic across network devices. 
As shown in Figure~\ref{fig:motivation}, the shuffling traffic is sent from mapper to reducer over a switch fabric, forming an aggregation-style data communication pattern.

\subsection{Programmable Switch}\label{subsec:prog_switch}
Traditional switching ASICs are designed with fixed functions. Under this property, adding a new feature requires designing and manufacturing a new ASIC, which incurs huge capital and engineering costs. However, programmable switching ASICs like Barefoot Tofino~\cite{switch-barefoot} and Cavium XPliant~\cite{cavium} make in-network computing~\cite{shah2018pcube} viable and deployable. They allow users to program the switch packet processing pipeline with a domain-specific language. The most well-known language to program the data plane is P4~\cite{p42014} and our P4COM is based on the P4 language. In detail, network administrators are able to (i) program the switch parser to identify user-defined packet formats (e.g., containing custom fields for keys and values), (ii) program the on-chip memory to store custom state (e.g., store keys and computed values), and (iii) program the switch tables to perform custom actions (e.g., write values from stateful memory to packets).

The Barefoot Tofino switch \cite{switch-barefoot} is a typical example of the new generation programmable switches. It consists of several pipelines. First, packets arriving at the front panel ports are parsed by parsers defined in data plane programs (written in P4 language). Then, the headers extracted by parsers and metadata are forwarded to an ingress pipeline. After a sequence of processing, packets go through a traffic manager and then are passed to an egress pipeline. Finally, after the egress pipeline, packets are forwarded to output ports. The processing in each pipeline is decomposed to multiple match-action tables.

Besides the stateless data (which is not preserved between packets, e.g., packet header, metadata), several stateful objects (which is preserved between packets, e.g., counters, meters, registers) are available in the Tofino switch. The most flexible stateful object is register, which is implemented using SRAM blocks and can be read/written by the data plane. Tofino switch also provides stateful arithmetic and logical units (SALUs) to perform simple computations on registers, metadata and constants. Network engineers can apply these operations to realize hash-based algorithms (e.g., bloom filters and sketches). Furthermore, the SALU modules also support additions and comparisons. This property is appealing to enable programmable switches to realize in-network computing.

\subsection{In-network Computation} \label{subsec:motivation}

\begin{figure}[t]
	\centering
    \includegraphics[width=3.3in]{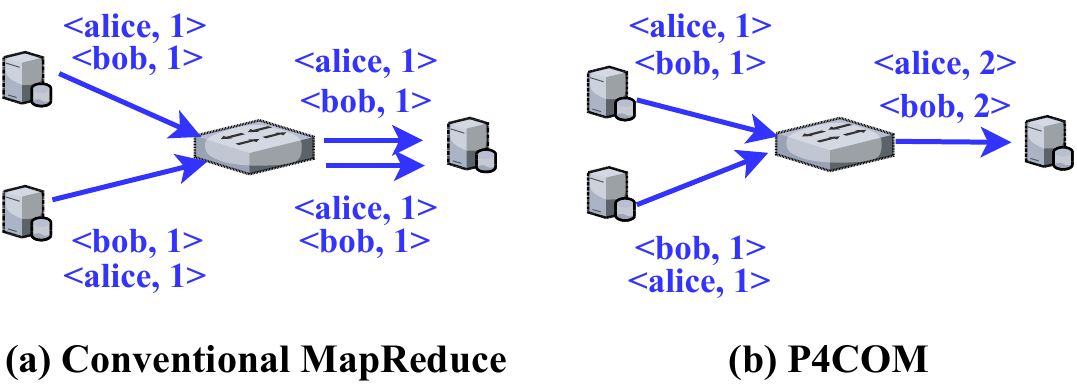}
	\caption{Word Count for conventional MapReduce and P4COM.}
	\label{fig:example}
    \vspace{-1em}
\end{figure}

We now use a typical Word Count example to illustrate the motivation behind in-network computation. In the MapReduce framework (Hadoop~\cite{hadoop}, Spark~\cite{spark, spark2012matei}, etc.), the Word Count benchmark runs as follows (see Figure~\ref{fig:example}):

\textbf{Conventional Procedure:} Source text files are located on different workers, and Map tasks are assigned to these workers. (i) Map tasks map each word (``alice'', ``bob'', etc.) in the text file to tuples ($<$``alice'', 1$>$, $<$``bob'',1$>$, etc.); (ii) These tuples are sent from mappers (servers running Map task) to reducers (servers running Reduce tasks) based on their hash values; (iii) Reduce tasks accumulate the total appearance of different words ($<$``alice'', 1$>$, $<$``bob'',1$>$, etc.). Each reducer is responsible for a subset of the entire vocabulary, and the final result is the combination of the results from all reducers.

\textbf{In-network Computation Procedure:} Before the $<$Word, Count$>$ pairs are passed to the NIC, workers serialize their own word lists into streams of packets, and send them to the data collector. When these packets arrive at the P4COM-enabled switch, they are parsed according to the pre-defined format and complete the reducer function. Temporary $<$Word, Count$>$ pairs (e.g., $<$``alice'',2$>$ $<$``bob'',2$>$, etc.) are stored inside the switch stateful memory. Later, mappers send data collection signal packets to reducers periodically. When signal packets arrive, the switch will assemble the $<$Word, Count$>$ pairs from the memory into the signal packets, and forward them to the reducers. At last, a final reducer is executed on the reducer node, dedicated for those ``unreduced'' pairs.

%There are some key differences between the conventional MapReduce framework and P4COM. First, in the P4COM data plane, the smallest unit of computation is packet, and each packet should have a fixed format for the switch to process. Second, the P4COM switch no longer simply forwards packets, and plays the role of intermediate reducer and store temporary $<$key, value$>$ pairs in the stateful memory. Finally, P4COM switch needs to assemble data packets from its memory and forward them to the final data collector.

\begin{table}[t]
	\centering
    \caption{Comparison of Existing Frameworks}
    \label{tb:Compare}

    \begin{tabular}{C{2cm} | C{1.5cm} C{1.1cm} C{1.1cm} C{1.1cm}}
		\hline
        System & Hardware Acceleration & User Friendly & Memory Efficient & Fault Tolerant \\
        \hline \hline
        Hadoop \cite{hadoop}  &  \xmark &  \cmark  &  \xmark & \cmark \\
        NetAgg \cite{netagg2014mai} &  \xmark &  \cmark & \xmark &  \cmark \\
        DAIET  \cite{sapio2017daiet} &  \cmark &  \xmark & \xmark & \xmark \\
        iSwitch \cite{li2019accelerating} &  \cmark &  \xmark & \xmark & \xmark \\
		SwitchML \cite{switchml} & \cmark &  \xmark & \cmark & \cmark \\
		ATP  \cite{atp} &  \cmark &  \xmark & \cmark & \cmark \\
        \hline
        \textit{\textbf{P4COM}}   &  \cmark &  \cmark &  \cmark  & \cmark \\
        \hline
	\end{tabular}
    \vspace{-1em}
\end{table}

\textbf{Theoretical Analysis of In-network Computation:} Offloading computation into the network can reduce the shuffling traffic volume. The shuffling traffic reduction is related to factors of key-value proportion, mapper-reducer ratio and data randomness. The key-value proportion is the packet valid payload size divided by total packet size, including header fields.
In our scenario, the Tofino switch handles a packet of MTU-size at most, when transferring a certain amount of data, it may generate more packets which incurs a relatively expensive header overhead. Assume dataset of size $D$ is uniformly distributed on each mapper server, with a mapper-reducer ratio of $N:1$,  the total bytes injected into the bottleneck link (last hop) in conventional manner is:
\[
    V = D * (1 + \frac{1}{R}) * N,
\]
while the bottleneck data volume for in-network aggregation scenario is:

\[
    V' = D * (1 + \frac{1}{R}) / \rho 
\]
where $R$ is the key-value proportion, and $\rho$ is the randomness factor.
Hence, the relative shuffling volume ratio is:
\[
   \frac{V'}{V} = 1 / (N * \rho)
\]
This reflects the fact that the data reduction ratio is correlated negatively to the data randomness.

\begin{figure*}[t]
    \centering
    \includegraphics[width=6.0in]
        {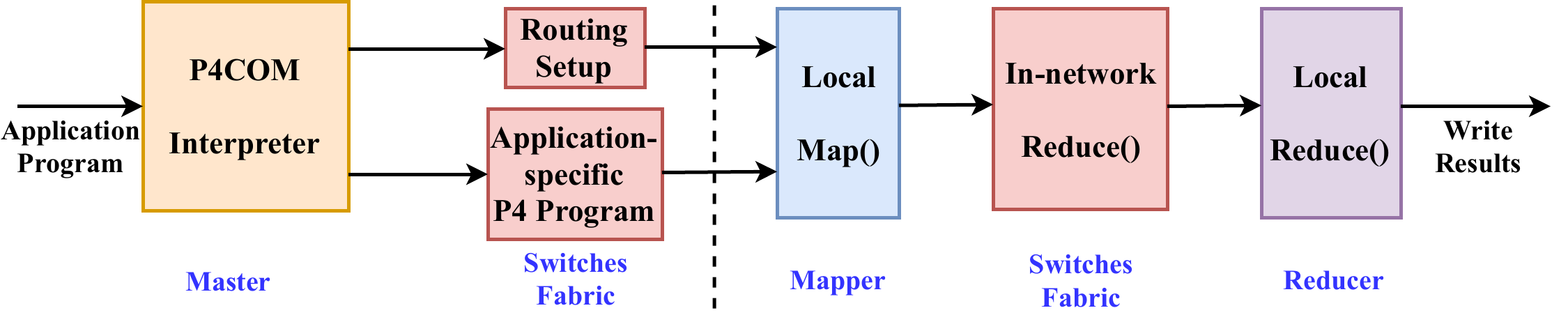}
    \caption{Workflow of P4COM. The dotted line separates configuration setup from computation data flow.}
    \label{fig:workflow}
    \vspace{-1em}
\end{figure*}

\textbf{Challenges of In-network Computation:} However, offloading computation tasks to switches is non-trivial and has the following challenges. First, the native switch interfaces (CLI and APIs) are primarily designed for network administrators to do network configurations and monitoring, which are not friendly for application developers to program computation tasks. Second, switches have limited on-chip memory for flow tables and packet buffers, while many computation tasks require a significant amount of memory. By performing computation tasks on switches, the precious switch memory resources would soon become a bottleneck. Third, aggregation traffic may suffer from packet loss in the network. If not being correctly handled , the packet loss would cause incorrect computation results in the switches.

There are some recent work proposing similar idea of in-network computation, such as DAIET~\cite{sapio2017daiet}, iSwitch~\cite{li2019accelerating}, SwitchML~\cite{switchml}, and ATP~\cite{atp}.  DAIET~\cite{sapio2017daiet} builds a functional prototype and is evaluated on Word Count examples. iSwitch~\cite{li2019accelerating} accelerates the training process of reinforcement learning applications based on NetFPGA. More recently, SwitchML~\cite{switchml} is proposed and considers many realistic issues, followed by ATP~\cite{atp} that aims for the more generic multi-job, multi-rack scenario. However, none of these proposals simultaneously meet the requirements of user friendly, memory efficiency, and fault tolerance (see Table~\ref{tb:Compare}). Therefore, we propose P4COM---an in-network computation framework that is user friendly, memory efficient, and fault tolerant.

\section{\sys Design} \label{sec:design}

P4COM is an end-to-end key-value pair computation framework that leverages programmable switches to reduce network traffic load. We assume the computation cluster is inside a modern data center and the network administrator has the complete control of all components in the switches and servers. We use a ToR to connect the master and workers as a basic architecture, and this can be extended to a multi-layer data center. We give a detailed description on P4COM components, protocols and data plane designs in this section.

\subsection{Overview}
P4COM targets at clusters consisting of end-hosts and programmable switch fabrics. The hosts in the cluster are labelled as master, mapper, or reducer according to their roles. The mappers and reducers are also known as worker nodes. The switch data plane is loaded with application-specific P4 program and executes in-network ``reduce()'' function.

Figure~\ref{fig:workflow} shows the workflow of a P4COM job. The whole P4COM job life time can be partitioned into two periods. The first period includes the compilation process and configurations setup. The compiler translates application tasks into P4 programs and allocate to target switches in the network. Application-specific routing schemes are configured after the compilation and loaded through control plane channel. They are responsible for forwarding shuffling data to correct in-network switches (``reduce()'' in the network). The second period is the execution of computation tasks in parallel. The mappers run local ``map()'' function to generate key-value pairs from data source. Intermediate data is shuffled to reducers by P4COM-specific TCP connections. During the shuffling, partial data is processed by ``reduce()'' on the switch data plane and temporary results are cached in the stateful memory units. Results stored in switch fabric will be forwarded to reducer periodically. Final results are collected by the reducers.

\subsection{Components}

We introduce the physical components of P4COM, namely master node,
worker node and switch fabric in this section.

\textbf{Master.} The master node is the central controller for the entire distributed application process. It is responsible for distributing mappers and reducers to worker nodes and delivering control plane messages.
In our P4COM testbed, the master also runs a final reducer process, as the collection node of final key-value pairs. A shim layer is running to interpret all received packets from the P4COM switch.
The master node initiates the computation workflow. Later, when intermediate data arrives, it extracts the keys and values from the packet stream. A last step reducer is executed to avoid duplicated keys in the end. When ``FIN'' signals are received from all reducers, the master terminates the process.

\textbf{Worker (Mappers \& Reducers).}
A distributed computation job includes multiple workers. In our prototype, each worker takes the role of a mapper or reducer. To take advantage of locality, mapper is assigned to nodes storing part of the data. Once receiving signals from the master, each mapper loads data from disk to memory and starts local ``map()'' processes to generate locally intermediate $<$key, value$>$ pairs. Before data is forwarded to the NIC, a shim layer is running to construct P4COM packets (see the packet header part below).

\textbf{Switch.} The programmable switch performs the intermediate processing for the $<$key, value$>$ pairs following Algorithm~\ref{alg:processpkt}. The switch executes application-specific P4 programs. It detects and parses P4COM formatted packets. If new data arrives, it applies operations on the keys and values. Intermediate results are stored in the stateful memory (register arrays). When collection signal packets arrive, the switch assembles the stateful results to the payload of the signal packets and forwards to the dedicated collectors.

\begin{algorithm}[!t]
	\scriptsize
	\DontPrintSemicolon  %Some LaTeX compilers require you to use \dontprintsemicolon instead
	\KwData{Intermediate key results $K\_array$; \newline
            Intermediate value results $V\_array$; \newline
            Memory usage counter $kv\_size$; \newline
            Mappers progress array $m\_states$; \newline
            Pre-defined memory bound $B$; \newline
            P4COM packet flags (CPA=Validity, CPK=ACK, CPD=Data).}
	\Begin{
		\tcc{Check validity of P4COM protocol.}
		\If{$pkt.CPA == 1$}{
            \tcc{Check validity for ACK packet.}
            \If{$pkt.CPK == 1$} {
                forward ACK packet to mapper\;
            }
			\tcc{Check validity for data packet.}
            \Else {
                forward $pkt.header$ to master\;
                \If{$pkt.CPD == 0 $}{
                    \If {$K\_array.exists(pkt.key)$}{
                    \tcc{Update value in the register.}
                    $V\_array.key  \longleftarrow pkt.operation(V\_array.key, pkt.value)$\;
                    }
                    \Else {
                        $V\_array.key  \longleftarrow  pkt.value$\;
                        increment $kv\_size$\;
                        \If {$kv\_size > B$}{
                            $pkt.CPD \longleftarrow 1$\;
                            recirculate packet to ingress\;
                        }
                    }
                }
                \If{$pkt.CPD == 1 $}{
                    $pkt.payload$ $\longleftarrow$ $kv\_size$\ $K\_array$\ $V\_array$\;
                    forward $pkt$ to master\;
                }
            }
        }
        \Else{
            forward normal TCP packet\;
        }
	}
    \caption{\sys ProcessDataPacket(pkt)}
    \label{alg:processpkt}
\end{algorithm}

\subsection{Protocols and Packet Format}
P4COM demands an L4 and application-layer ensembled protocol to accomplish the computation process and deal with packet losses.
Similar to other MapReduce frameworks, P4COM supports commutative and associative operations. For the applications in our prototype, we utilize these basic operations that are available in the P4-supported commodity switch, namely MAX, MIN and ADDITION. 
In order to achieve reliability, P4COM uses a specialized TCP stack for both data streams and control message streams. The P4COM fields are appended to the P4COM-enabled TCP header.
A set of TCP ports are reserved for P4COM. P4COM switches use these ports to invoke the application-specific packet processing logic for P4COM queries.
Other non-P4COM switches do not have to understand the P4COM packet format. They treat P4COM related packets as normal packets.

\begin{figure}[t]
	\centering\includegraphics[width=3.0in]
        {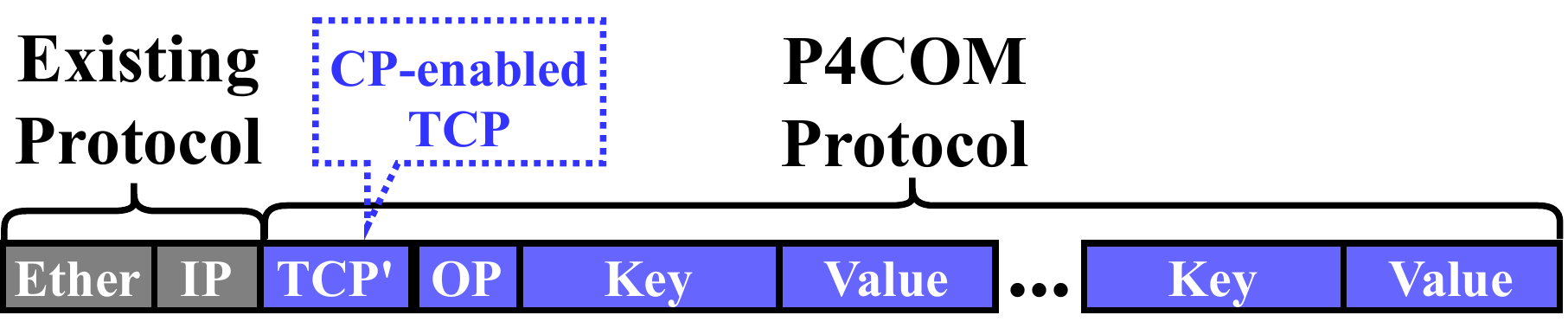}
	\caption{P4COM Packet Format}
	\label{fig:whole_packet_header}
	\vspace{-1em}
\end{figure}

P4COM has three types of packets, i.e., master control message, mapper data stream, and reducer data collection signal. 
Master control messages deliver commands and scripts to switches and servers, and these packets use the normal TCP format. 
Worker data streams are the packets carrying $<$key, value$>$ pairs after local computations in the mapper nodes. 
We show the format of data packets in Figure~\ref{fig:whole_packet_header}. There is no modification in L2 and L3. In L4, we apply a special cutting-payload-enabled TCP header (see the fault tolerance part below). As for the payload, we treat it as part of headers and extract them in the parser logic of P4. The first 8-bit payload is reserved for the operation (OP) field. The available OP options include MAX, MIN and ADDITION. This field informs the switch of the desired operation to execute on following key-value pairs. After the operation field, we ensemble the list of key-value pairs. For simplicity, we use 64 bytes for each key and 4 bytes for each value. The data packet can get extended to the size of MTU, which is able to carry 20 key-value pairs.

\subsection{Challenges and Design Choices}

There are some technical challenges to realize P4COM with programmable switches and commodity servers.

Firstly, the domain-specific language to program the switch data plane is P4~\cite{p42014}. The P4 language is primarily designed for network administrators to define parsers and match-action tables in the pipelines. The straightforward applications are network monitoring and configuration setup. However, it is not friendly to developers who express their own computation logic, especially the ``reduce()'' function.

Secondly, according to our protocol design, the intermediate data of $<$key, value$>$ pairs is stored inside the stateful memory. However, the memory resource (SRAM) to store stateful variables on Barefoot Tofino Switch is limited to several hundreds of Megabytes~\cite{switch-barefoot}. While this capacity can grow with more engineering efforts, it is still insufficient to meet the demand in distributed applications. For an application with a demand more than the limit of switch memory, we will suffer from memory overflow problem. Hence, it is impossible to store all intermediate results in the switch memory.

Thirdly, our framework has a strict requirement for packet reliability. The $<$key, value$>$ pairs inside the data packets are retrieved once arriving at the P4COM switch. To save bandwidth, a straightforward design is to drop the packet when its data is retrieved. However, it makes receivers unable to distinguish packet losses from normal retrieved packets.
Furthermore, when packet drops occur, the default loss recovery mechanism in TCP may lead to incorrect result (e.g., duplicate operations for key-value pairs in the dropped packets).

To address these challenges, P4COM incorporates a micro compiler to translate application logic to P4 programs, an efficient memory management approach, and a reliability control protocol. We now describe each one of them below.

\textbf{Translating application logic to P4.}
To facilitate the usage, we develop a programming framework for P4COM, which exposes an API for atomic operations to programmers. Programmers are assumed to be the network administrator who has complete control over all switches and servers. Programmers need to write a P4COM program consisting of the definitions of several atomic operations.

We develop a lightweight interpreter to compile the programmer-defined P4COM program. In this user-defined P4COM program, one atomic operation is dataset path setting, which assigns the IP and file path to a variable. Another operation is mapper assignment, which allocates mappers and local reducers to dedicated datasets. The last operation is reducer assignment, which allocate reducer to dedicated mappers. The compiler parses the abstract syntax tree of the program, and then uses a greedy algorithm to allocate intermediate reducers to programmable switches. The greedy objective of reducer allocation is to minimize the sum of distance between mappers and reducers. By allocating intermediate reducers to different switches, application-specific routing scheme is necessary. The compiler will also setup the routing policy for each many-to-one mapper-reducer pattern.

We provide a compile-and-run python script to launch the compilation and collect job progress information. With our P4COM micro compiler, developers can focus on their application-specific MapReduce logic without directly managing the data plane level functions.

\textbf{Memory management.}
To address the limited memory challenge, we first leverage the fixed key space to approximate the memory occupancy on switches. We construct a heap on each mapper side, merging new keys when new packets are assembled. We set a threshold of key limit (100K in testbed) as an upper bound. This heap is used to approximate with the switch memory occupancy, with the observation that the merged key-value pairs memory is proportional to the key-only memory. Whenever the heap reaches the preset upper bound, we trigger a collection signal to the P4COM switch to push current results to the master node. At the same time, we clean the heap on the mapper and the memory on switch.
This is similar to a batch algorithm in a high-level view, as we process the data one block after another.

However, as the number of mappers grows, this mechanism alone may suffer from the synchronization problem and introduce extra latency. To avoid the memory overflow problem, we propose our dynamic memory management with an extra active array inside the P4COM switch. When a new data packet arrives at the P4COM switch, the data is extracted by our parsers. For each $<$key, value$>$ from this packet, if the key hits one field in the KV arrays for intermediate results, we update its corresponding value field after the operation. In this case, no extra memory unit is consumed. Otherwise, if a key is new, we apply several hash functions to it in order to locate a new block for this $<$key, value$>$ pair storage. At the same time, we will increment a ``global'' memory counter, which indicates the number total distinct keys currently stored. If the counter value exceeds a certain threshold, we will modify the CP header field inside data packet, ensemble temporary KV arrays and forward to master node. When master node receives this packet, it will inform mappers to lower their sending rates. In other cases, the sending rate of the mappers follow traditional TCP congestion control algorithms, by default TCP-Cubic for Linux. With this strategy, the memory usage of programmable switch can be bounded to a reasonable limit.

\textbf{Fault tolerance.}
Packet losses are inevitable in datacenter network due to buffer overflow, transmission error, etc. To distinguish undesired packet losses from intended P4COM packet drops, inspired by the cutting payload (CP) method~\cite{peng2014CP}, we develop our application-specific CP mechanism. Specifically, when data packets arrive at the switch, the parsers will extract each field including the flags in L2, L3, L4, and application-specific ``payload''.
After operations on the latest data and updates of switch memory, the switch will modify some flags in the CP header if state transition occurs. For example, if the memory usage is above the preset limit, the CPD flag will be set to inform the master.

We also need a mechanism to detect and react to packet losses and retransmissions. At the initialization step, our P4COM switch will allocate a register array (size equals to the number of mappers) as per-flow counters. This array is responsible for maintaining the progress states of each mapper. The values of the array are initialized to ACK sequence number of the first data packet from each mapper. When a data packet comes, the switch first compares the current ACK sequence number with the register value associated with this mapper. If the ACK is no larger than the corresponding register value, it implies that data in the packet has been processed before, and the switch will simply drop this packet. Otherwise when related operations are completed, the switch will increment the register value of the mapper by one. Hence, when duplicated packet arrives, the intermediate results will not be modified. P4 switch has an option to recirculate packet from egress to ingress. By exchanging the source and destination field of the retrieved data packet, we can generate an ``ACK'' packet and push back to the sender. To trace the packet sequence, it is necessary to add a monotonically increasing integer. If the sender receives a non-continuous integer, then the previous packet is supposed to be lost, and a replicated send action will be triggered on the mapper side.
\section{Practical Considerations} \label{sec:discussion}

\subsection{Issues of In-network Computation}
The common concerns for offloading computation to the switch data plane include performance degradation and resource limitation. The Barefoot Tofino Switch provides up to 6.5 Tbps throughput, and in most cases the core network bandwidth is abundant to support distributed computation \cite{roy2015inside, singh2015jupiter}.
In general, modern switches are equipped with tens of megabytes of on-chip memory. Since many key-value pairs are small values (e.g., $<$100-byte values for 76\% read queries in the Memcached deployment at Facebook \cite{memcache2014fb}), the switch on-chip memory is able to store hundreds of thousands of items for computation, leaving enough switch resources for conventional network functionalities. With our memory management scheme, the SRAM resource utilization in the switch fabric can be bounded to a constant. For large key-value pairs that do not fit in one packet, one can always divide an item into smaller chunks and retrieve them with multiple packets. Note that multiple packets would always be necessary to collect all the intermediate values stored in the switch stateful memory.

\subsection{Compatibility with Existing Frameworks}
Programmers are unwilling to apply new frameworks if they have to
change all programs from scratch. P4COM employs an L4 plus application
layer ensembled protocol. In our prototype, we have implemented an
interpreter to interact with a lightweight MapReduce framework. A shim
layer parses the key-value pairs from the first stage mappers and
ensemble to P4COM packets. In our prototype, users can apply
our shim layer as a plugin to their programs, without
modifying original versions. However, to be compatible with all existing
frameworks requires more engineering efforts to parse their
data structures. This extension is feasible in the technical
aspect, but we leave it to future work if more users are interested
to apply our P4COM framework.

\subsection{Limit of Use Cases} 
P4COM is a prototype to demonstrate the feasibility of in-network computing. Due to resource limitations, all evaluations are carried out in either the Mininet simulation environment or a small testbed consisting of one Tofino switch, several commodity servers and 10Gbps links. A capacity test on a large-scale deployment is left for industry interests and future works. As for the type of MapReduce applications supported, we want to reach the limit of data plane SALU. Currently, only fixed-bit additions and comparisons are supported. Any MapReduce job that can be interpreted in this form of computation can benefit from P4COM. 
%As the data plane functions are evolving rapidly, we will have more operations available to our framework.

\section{Evaluation} \label{sec:evaluation}

We use a combination of Mininet simulations and testbed experiments to evaluate the performance of P4COM. Specifically, we aim to answer the following questions:
\begin{itemize}
    \item \textbf{Can P4COM accelerate computational tasks?} We evaluate P4COM under various workloads and network scenarios, running applications including Word Count and Deep Neural Network (DNN) training, etc.
    \item \textbf{How well does P4COM mitigate Incast?} This is a straightforward benefit from our framework. By caching partial intermediate results in the switch memory, we significantly decrease the possibility of Incast.
    \item \textbf{Can P4COM work well in practice?} We build a prototype consisting of a Barefoot Tofino switch and several hosts with 10G NICs, validating that P4COM achieves near line-rate throughput and handles packet losses efficiently.
\end{itemize}

\subsection{Implementation and Dataset}
\textbf{Implementation.}
The P4COM implementation consists of controller logic, switch data plane, an interpreter on master and worker servers to interact with application data. The master distributes initial setup scripts for workers and control the computation process by constantly probing the data collection signal to the switch. Workers execute local MapReduce tasks, and construct P4COM packets with a shim layer before NIC.

In the Mininet simulations, the controller is written in Python and resides on the master node. In the testbed environment, the controller is written in specific shell scripts provided by Tofino Shell and is located in the OS of Tofino switch.
Both controllers communicate with their targets through Thrift APIs at runtime. The controllers are able to update table rules, modify port configurations and monitor stateful memory.

We also implement a set of lightweight distributed MapReduce APIs in C++. It focuses on the MapReduce logic instead of the management and monitoring functions. Hence, it achieves better performance for our specific input formats compared with existing softwares (e.g., Apache Hadoop~\cite{hadoop} and Spark~\cite{spark, spark2012matei}). We apply these APIs on mapper and reducer nodes as local computation modules. Above the APIs, a Python shim layer performs translation between the local results of MapReduce program and the P4COM packets.

\textbf{Dataset.}
For Word Count, we download several public datasets and analyze their data statistics. First, there is a Wikipedia Dumps containing 5105919 distinct keys. Second, we collect a Web Link graph data with 875713 nodes and 5105039 edges. Finally, we obtain tcpdump file from an enterprise datacenter network with 400 million packets and 20 million 5-tuple flows. 

For DNN training, we refer to several recent popular models~\cite{he2016deep, simonyan2014very, ronneberger2015u} in Image Processing domain. The statistics are shown in Table \ref{tb:DNNs}.
As for DNN training, we map the index of each model tensor to an incremental key, and then quantize the tensor cell value to match our P4COM value field. A DNN training process can be decomposed to multiple rounds of MapReduce, with the `reduce()' function set to SUM.

\begin{figure}[t]
	\centering\includegraphics[width=3.0in]
        {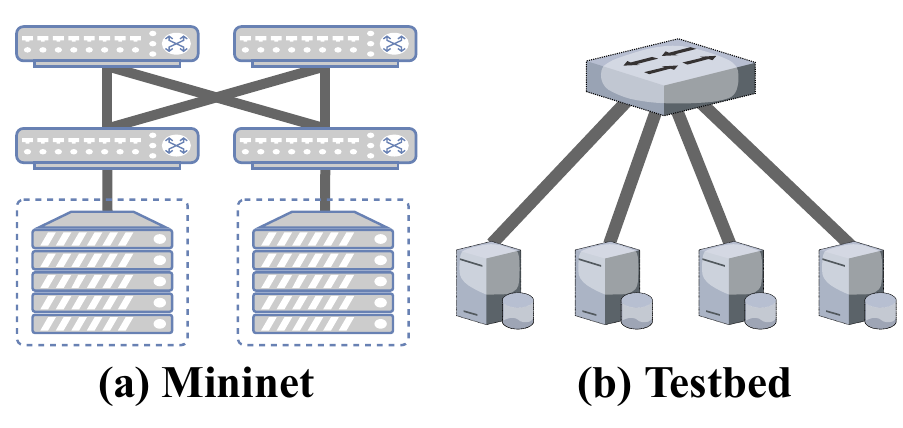}
	\caption{Topology. Mininet consists of behavior model software switches and virtual hosts (both surrounded by dotted lines). Testbed contains one Tofino switch and multiple commodity servers. }
	\label{fig:topo}
	\vspace{-1em}
\end{figure}

\subsection{Mininet Simulation}
\textbf{Setup.} The simulation prototype is implemented in Mininet on top of a physical server. The base operation system is Ubuntu 14.04, with a version 3.13.0 Linux kernel. The server is equipped with 24-core CPU and total 176 GB memory. 

The simulation topology is shown in Figure~\ref{fig:topo}~(a).
We use a two-tier network topology consisting of 4 switches (2 ToRs and 2 Aggregations) and 12 hosts (6 hosts under each rack).  The simulated link rates are  1Gbps between switches. We randomly split each of our datasets into partitions of equal size.  Partitions of datasets are assigned to hosts logically in our tests. 
%The base operating system serves the role of controller, and sets up the configurations of switches. 
On mappers, there are local reducers producing first stage ``reduce()'' to minimize the intermediate traffic.

\begin{table*}[htbp]
	\centering
	\caption{6:1 mapper-reducer DNN Models training Results}
    \label{tb:DNNs}	
	
    \begin{tabular}{c c R{1.5cm} R{1.5cm} R{1.5cm} R{3cm} R{1.5cm} R{2.5cm}}
		\hline
        Model & Dataset & \# Training Parameters & Epochs & Gradient Vectors & Quality Metric & Baseline Quality & P4COM Transmission Time Reduction Rate    \\
        \hline \hline
         ResNet50 \cite{he2016deep}  &  ImageNet  &  25,559,081 &  90 & 161 & Top-1 Accuracy & 75.37\% & $66.3\% \pm 5\%$ \\
         ResNet20 \cite{he2016deep} &  CIFAR10  &  269,467  &  328 & 51 & Top-1 Accuracy & 90.86\% & $62.72\% \pm 3\%$ \\
         VGG16 \cite{simonyan2014very} &  CIFAR10  &  14,982,987  &  328 & 30 & Top-1 Accuracy & 86.32\% & $63.5\% \pm 5\%$ \\
         U-Net \cite{ronneberger2015u} &  DAGM2007 &  1,850,305 & 2,500 & 46 & Intersection over Union & 96.4\% & $64.29\% \pm 5\%$ \\
         \hline
	\end{tabular}
    \vspace{-2.5em}
\end{table*}

\textbf{Schemes Compared.}
To show the performance gain of P4COM, we compare several scenarios of the many-to-one (number of mappers versus number of reducers) pattern.
In all the tests, we implement a list of three-to-one, six-to-one and twelve-to-one mapper-reducer ratio. In each test, each mapper takes a random partition of the dataset.

\begin{itemize}
    \item \textbf{Hadoop~\cite{hadoop}.} This is the baseline implemented by our lightweight distributed Hadoop framework. It aims to provide better performance compared with popular open source version.
    \item \textbf{DAIET~\cite{sapio2017daiet}.} This is the result from DAIET, which aggregates traffic in the network but without consideration of the memory efficiency and fault tolerance.
   \item \textbf{P4COM.} This scheme includes all components of P4COM and implement a sorting algorithm on intermediate keys before the results are sent to the NIC. This is the general performance indicator of P4COM.
\end{itemize}

\textbf{Scalability.}
Scalability is a key concern for any distributed computing framework. In our simulations, we provide a user-friendly interface to show how to scale P4COM to a multi-rack environment. The scripts are written in Python, interacting with the topology setting module in Mininet.

After running several jobs of multiple mappers plus multiple reducers within P4COM Mininet prototype, we find that all modules work fluently, including the core intermediate switch processing and routing. Results show that the final results of all three applications are coherent with the ones produced by conventional server-centric approaches. The routing scheme correctly distributes traffic evenly to different reducers. This proves the logical correctness of our central controller. 

\begin{figure}
    \subfigure[WikiReduce]{
    	\includegraphics[scale=0.325]
            {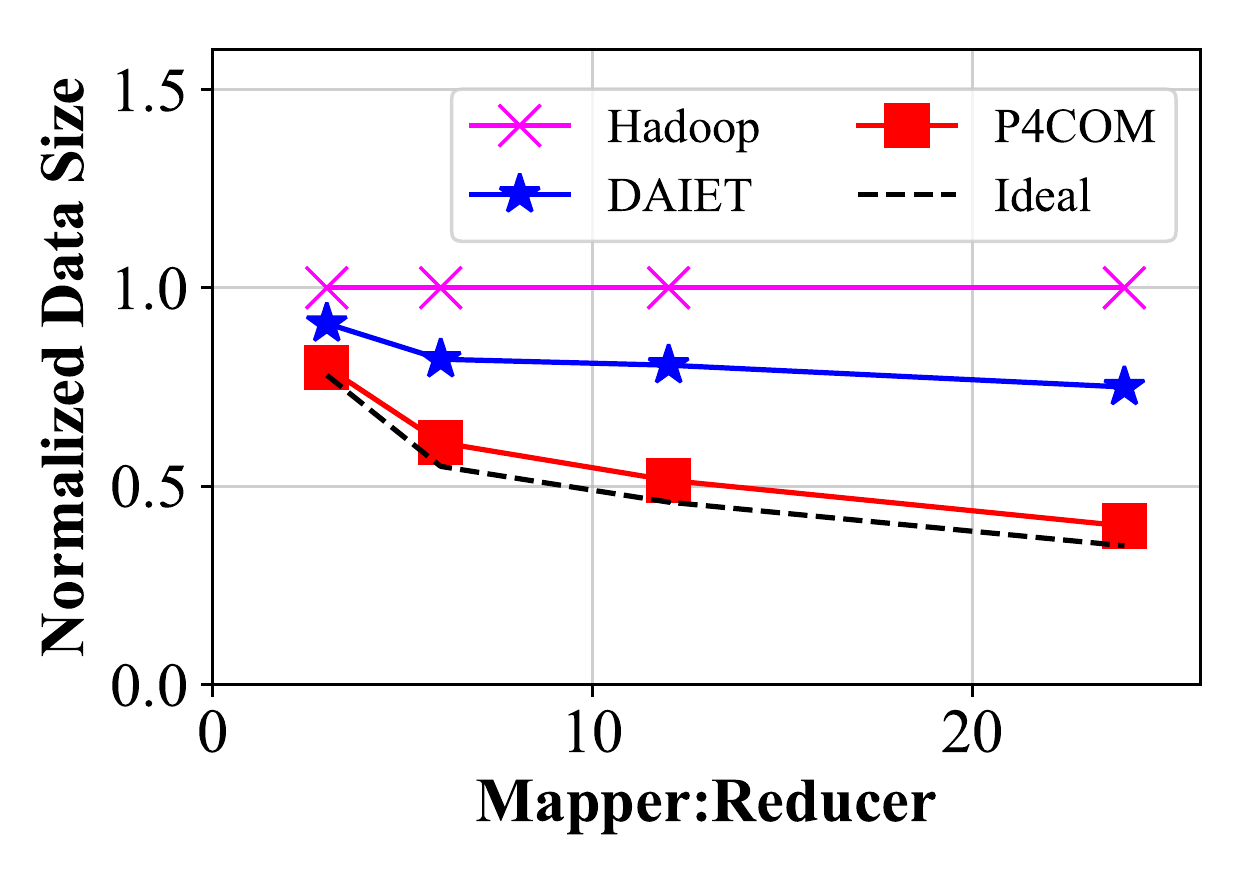}
	}
    \subfigure[ResNet20]{
    	\includegraphics[scale=0.325]
            {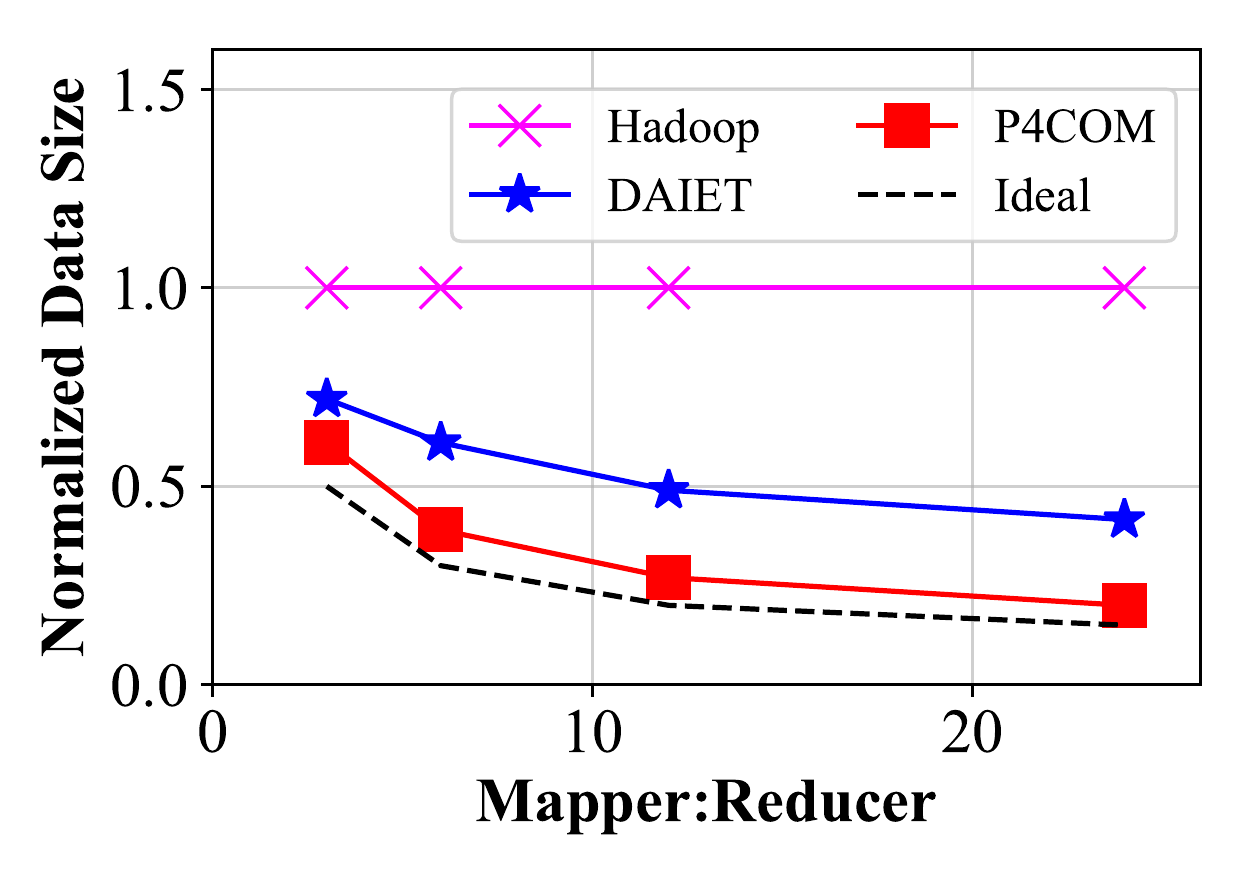}
	}
    \vspace{-1em}
    \caption{Mininet Results on Shuffling Traffic Volume Reduction. The Y-axis is the normarlized traffic size over Hadoop's on the last hop.}
    \label{fig:shuffle_reduce}
    \vspace{-1em}
\end{figure}

\textbf{Incast Mitigation.}
One advantage of P4COM over existing server-centric frameworks is the reduction on shuffling traffic size, which often leads to the Incast problem without careful handling. We evaluate the traffic reduction ratio on different datasets in our P4COM Mininet prototype. The results are shown in Figure~\ref{fig:shuffle_reduce}. The Y-axis implements a normalization procedure by dividing the intermediate traffic size measured in Hadoop scenario. The Ideal case is obtained based on \S\ref{subsec:motivation} assuming uniform data distribution and zero loss.

Our observation is that the P4COM achieves highest shuffling data reduction rate in both WikiReduce and DNN training scenarios. As the number of mappers increase, the data reduction rate becomes greater. Table~\ref{tb:DNNs} shows some sample results running DNN training jobs on P4COM. The average shuffling data reduction rate with 6 mappers work simultaneously is above 60\%. Notice that the DAIET may lead to less reduction rates for the same mapper-reducer ratio compared with P4COM, due to frequent out of memory losses and trigger retransmissions frequently.

\begin{figure}
    \subfigure[WikiReduce]{
    	\includegraphics[scale=0.325]
            {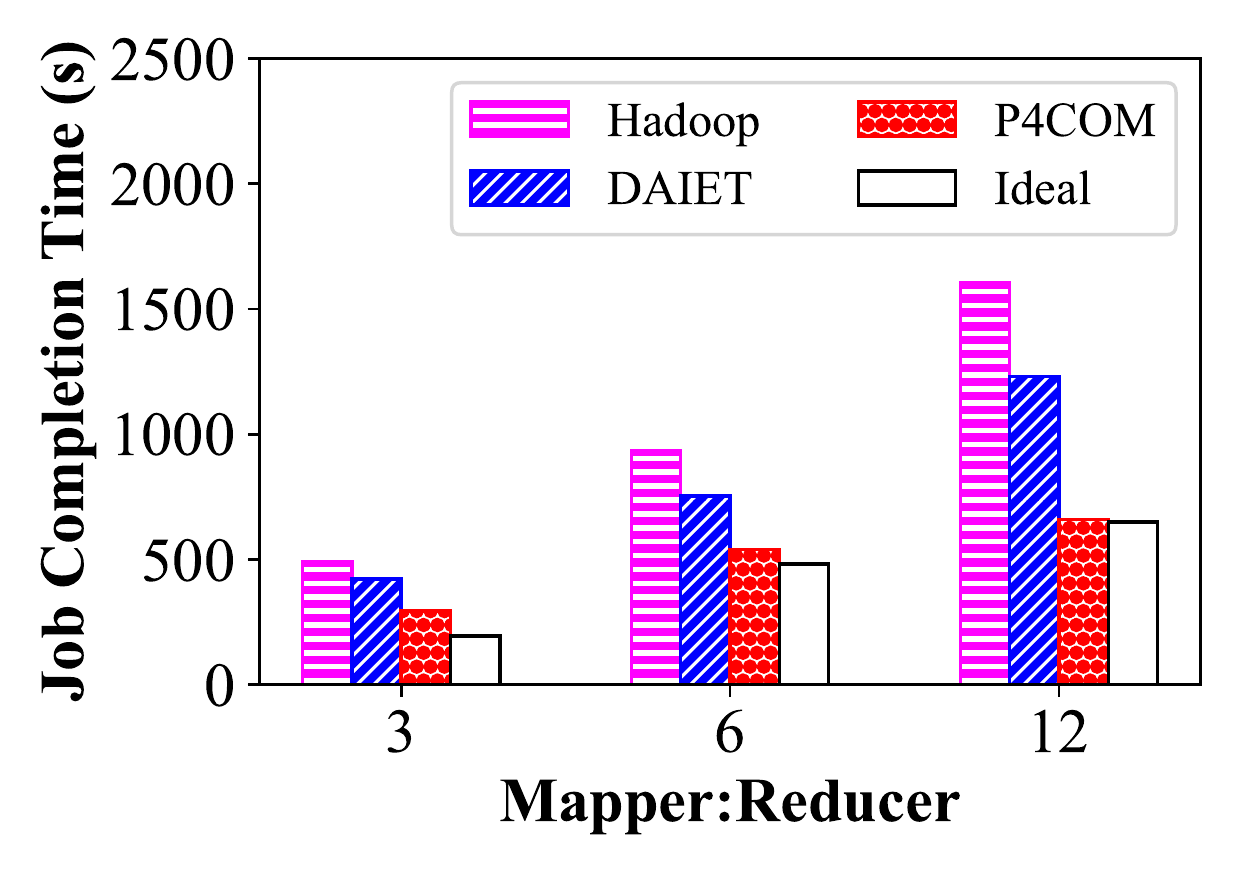}
	}
    \subfigure[ResNet20]{
    	\includegraphics[scale=0.325]
            {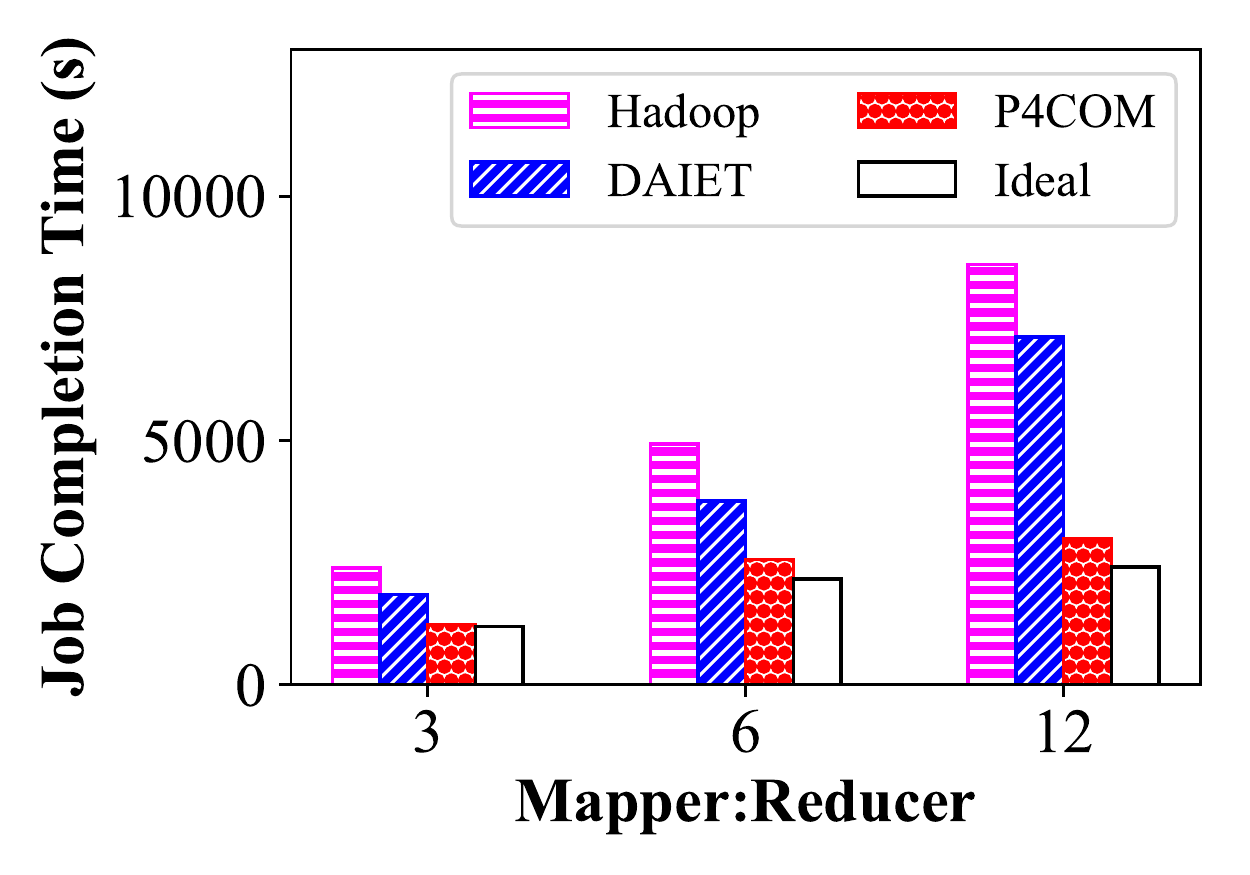}
	}
    \vspace{-1em}
    \caption{Mininet Results on Job Completion Time under 1Gbps network.}
    \label{fig:shuffle_fct}
    \vspace{-1em}
\end{figure}

Figure \ref{fig:shuffle_fct} shows the Job Completion Time (JCT) variation according to various mapper-reducer ratio. It demonstrates that our memory management and fault tolerance mechanism can significantly reduce the shuffling process, thus leading to much smaller JCTs. We also observe a sharp increase of packet losses when increasing the mapper-reducer ratio for DAIET.

\begin{figure*}[ht]
\centering
    \subfigure[WikiReduce]{
    	\includegraphics[scale=0.4]
            {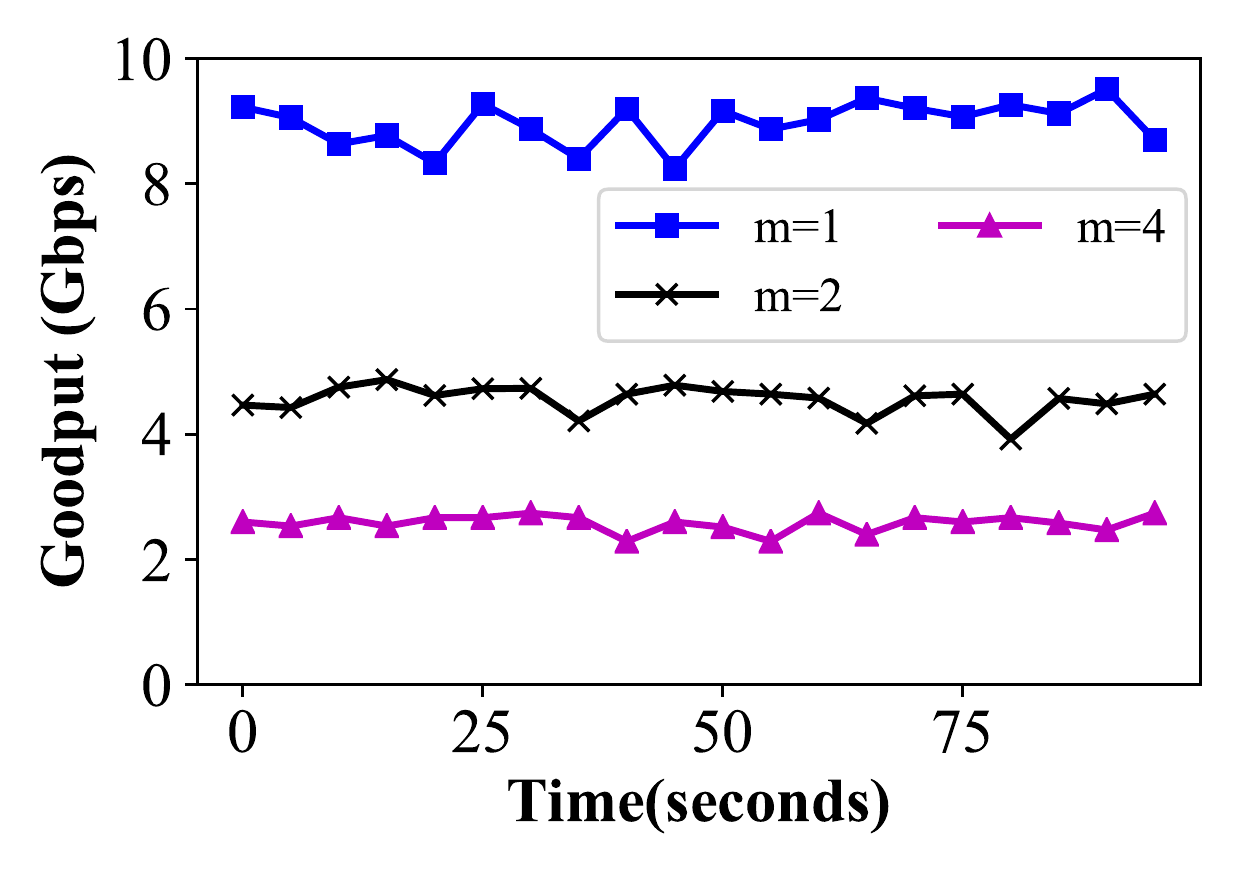}
    }
    \subfigure[RankQuery]{
    	\includegraphics[scale=0.4]
            {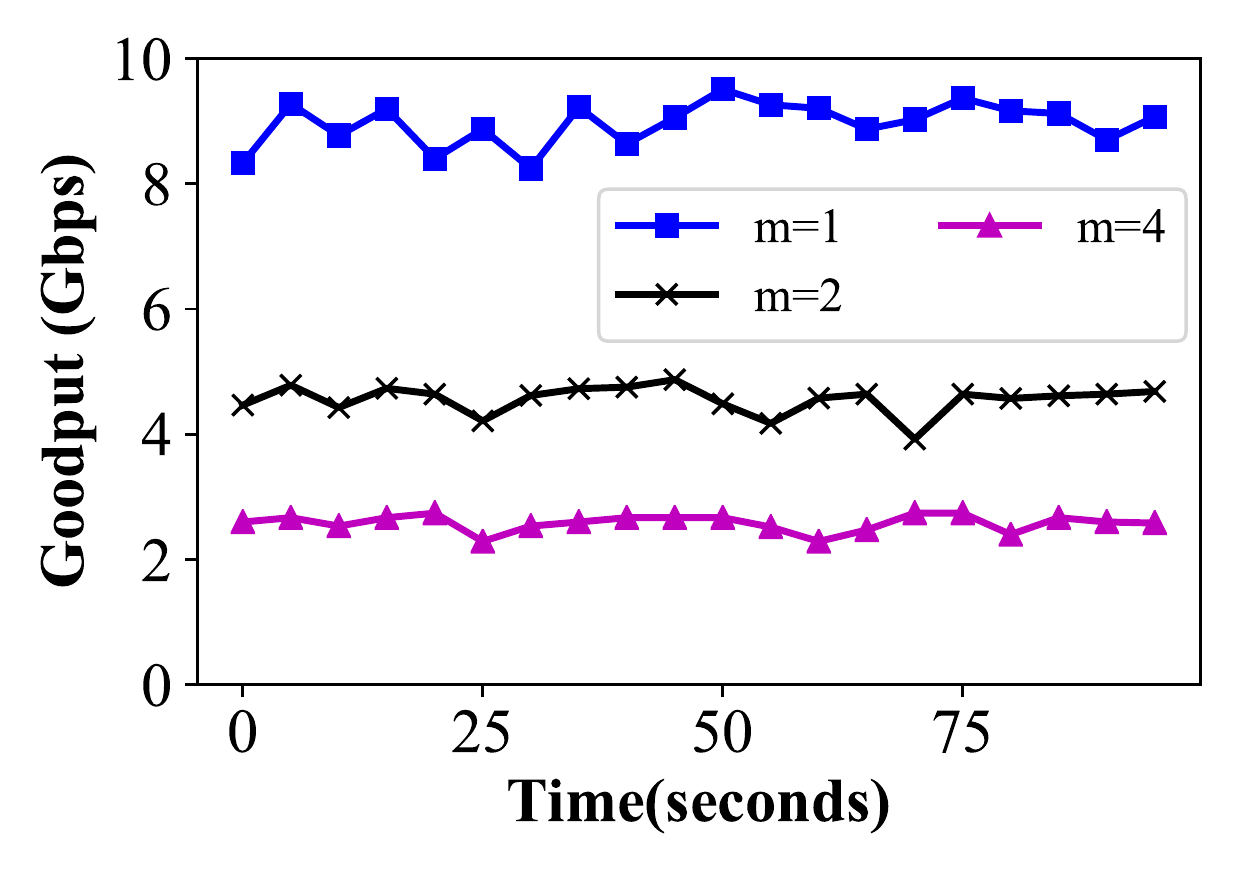}
    }
    \subfigure[FlowStatistics]{
    	\includegraphics[scale=0.4]
            {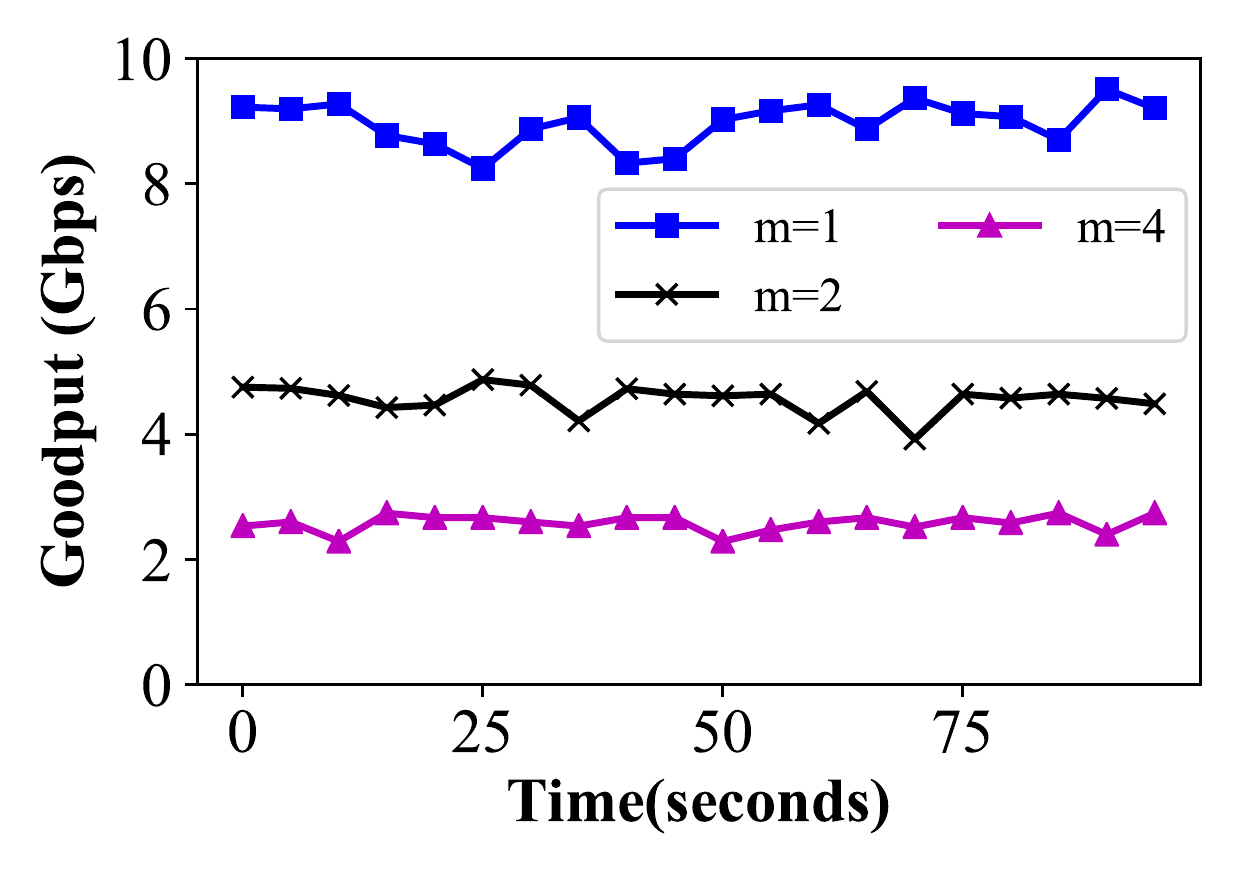}
	}
    \vspace{-0.5em}
    \caption{Average Goodput (one measurement on volume of application
    data successfully transmitted every 5 seconds). $m$ is the number
    of mappers per worker.}
    \label{fig:tput}
    \vspace{-1em}
\end{figure*}

In summary, the shuffling data size reduction is related to data randomness and number of mappers. For Hadoop, the shuffling traffic grows linearly as the number of mappers increases. For DAIET, inefficient memory usage and frequent losses severely degrade its overall performance. Compared to DAIET and Hadoop, P4COM achieves much higher shuffling data reduction ratio and slower JCT increase, when increasing the mapper-reducer ratio. Based on our Mininet results, programmers can setup their own ratio of mappers and reducers according to the application requirements.

\subsection{Testbed Experiment}
\textbf{Setup.}
We have built a P4COM prototype on a testbed which consists of one 3.2Tbps Barefoot Tofino switch and four Huawei RH1288 V3 servers. All links between the switch and servers have the same capacity of 10Gbps. The Tofino switch has a total of 32 front panel ports. Each port is associated with a SerDes block of four lanes with a maximum throughput of 100Gbps. Each server has 24 Intel Xeon E5-2670 CPU cores and 128GB (8 x 16GB DRAM) memory. The operating system on the servers is Ubuntu 16.04.1, with the 4.10.0 Linux Kernel.

The topology of our testbed is shown in Figure \ref{fig:topo} (b). We assign one server as master node, which distributes all computation tasks to the worker nodes. The baseline MapReduce APIs are the same as we described above. This implementation is lightweight and efficient in terms of core MapReduce logic. The shim layer is activated on each worker node. It is responsible for parsing intermediate results from local mappers and pass the data to our P4COM sockets.

\textbf{Goodput Measurement.}
We measure the average goodput (application data transmission rate) of P4COM jobs in each mapper process. The granularity of our goodput measure is 5 seconds. We show the average goodput variation within a period of 100 seconds in Figure \ref{fig:tput}. We use $m$ to denote the number of mappers on each server. $m=1$ gives a mapper-reducer ratio of 3:1. $m=2$ means 6:1 and $m=4$ refers to 12:1. For WikiReduce benchmark, the average goodput within the whole period is 9.13 Gbps when $m=1$. Increasing $m$ to 2 and 4, we obtain 4.26 Gbps and 2.31 Gbps respectively. Similar improvements are obtained for Flows Size application and Rank Query.
% For Flows Size application, the average goodput under three settings
% are 9.22 Gbps, 4.25 Gbps and 2.35 Gbps respectively.
% For Rank Query application, the average goodput are 9.19 Gbps, 4. 36 Gbps
% and 2.33 Gbps respectively for $m=1,2,4$ respectively.
Our P4COM prototype implements the application-level fault-tolerance polity on top of TCP-Cubic~\cite{ha2008cubic}. Due to the additive-increase multiplicative-decrease (AIMD) feature of TCP-Cubic, we observe temporary goodput variance. Nevertheless, our P4COM prototype achieves near line-rate goodput throughout all test cases.
\section{Related Work} \label{sec:related}
In this section, we summarize recent research works on data center networks (DCNs), programmable switches and prior efforts to utilize in-network computation (INC).

\textbf{Data Center Network (DCN).}
Research on data center network (DCN) is popular in recent years.
Different from the Internet, the administrators of DCNs have complete
control over all hardware devices.
Some propose new network architectures~\cite{al2008fattree, vl2009greenberg, osa, roy2015inside, singh2015jupiter} to support fast growing traffic demand. 
Some apply scheduling policies~\cite{varys2014mosharaf, bai2015pias, coda2016hong, karuna, bytescheduler, dlcp} to minimize flow/coflow completion times.
Some propose novel transport protocols~\cite{alizadeh2010dctcp, vamanan2012d2tcp, zhu2015dcqcn, ear, hpcc, mlt, gemini, transport-survey, aeolus} to achieve high bandwidth and low latency in DCN. 
Distributed computation services produce significant network traffic inside modern DCNs. To address it, some resource management frameworks~\cite{yu2009distributed, ahmad2014shufflewatcher} are proposed. However, such frameworks will add high complexity and cost overhead to the whole network management.

\textbf{Programmable Switch.}
There is a lot of research work leveraging switch programmability to improve networks or systems in various domains, such as in-network telemetry (INT)~\cite{kim2015band, hpcc, pint}, distributed in-network caching~\cite{li2016SwitchKV, jin2017netcache, distcache}, consistency guarantee~\cite{jin2018netchain}, load balancing~\cite{katta2016hula, miao2017silkroad, tiara}, and packet scheduling and fair queueing~\cite{pifo, pieo, sharma2018approximating, calenderq, hcsfq}.
Specifically, HPCC~\cite{hpcc} is the seminal INT-based high-performance congestion control for the DCN; SwitchKV~\cite{li2016SwitchKV} put a cache lookup table into switches with SDN-enabled content based routing, which significantly reduces the average cache latency; PIFO~\cite{pifo} is one of the earliest work on hardware packet scheduling and fair queueing, inspiring a bunch of work~\cite{pieo, sharma2018approximating, calenderq, hcsfq} in this line. Besides, there are many works on facilitating the programmable switch development, such as PISCES~\cite{shahbaz2016PISCES} that supports P4 prototyping on software, and various verification systems~\cite{liu2018p4v, vera, aquila} for programmable networks.

\textbf{In-Network Computation (INC).}
The idea of in-network computation (INC)~\cite{wetherall1999active} has emerged since two decades ago, but lacks attention due to hardware inflexibility until the recent emergence of commodity programmable switches~\cite{p4mr, sapio2017daiet, li2019accelerating, switchml, atp}. 
DAIET~\cite{sapio2017daiet} revisits several ideas on  the opportunity of in-network computation, with a preliminary system design and implementation. 
iSwitch~\cite{li2019accelerating} builds an in-switch computing framework for deep reinforcement learning applications based on NetFPGA. More recently, SwitchML~\cite{switchml} is proposed and considers many realistic issues, followed by ATP~\cite{atp} that aims for the more generic multi-job, multi-rack scenario.
However, all these in-network processing frameworks are not user friendly, and have many other insufficiencies (Table~\ref{tb:Compare}).
%Therefore, we propose P4COM to simultaneously achieve user friendly, memory efficient, and fault tolerant.
\section{Conclusion} \label{sec:conclusion}

Motivated by the capability and flexibility of modern programmable switches, we propose to offload some computational tasks to the data plane. We design P4COM, an end-to-end framework to enable in-network computing (INC). P4COM provides a user-friendly interpreter to translate application logic into data plane programs, along with an efficient memory management plus a fault-tolerance policy. 
Mininet simulations and testbed results show that P4COM can reduce the shuffling traffic significantly, and work at line rate under both 1Gbps and 10Gbps networks.
For the future work, we are considering to extend the scale of P4COM to the realistic multi-job, multi-rack data center network (DCN) scenario. 
%Besides, we will search for more efficient loss-recovery mechanisms to tackle performance challenges under 40Gbps and 100Gbps network.

% Future directions include:
% HDFS compatibility ( Enable P4COM to interact with distributed databases. )
% Multi-job Scheduling ( Given multiple
%       applications, a scheduling algorithm to optimize resource usage
%       is necessary. ),
% Dynamic Job Arrivals ( Dynamically arriving jobs may
%       require incremental compilation or virtualization mechanisms. ).
%, which we intend to explore further.

%   Currently P4COM only support pre-defined
% user program before the network starts.
%      When the simulated network is running, P4COM cannot change the
%      functions on the switch.
      %Supporting

%\clearpage

\bibliographystyle{IEEEtran}
\balance
\bibliography{IEEEabrv,p4com}

\end{document}